\documentclass[amsmath,aps,pra,twocolumn]{revtex4-1}
\usepackage{graphicx}
\usepackage{times}
\usepackage{slashed}
\usepackage{bm}
\usepackage{natbib}
\usepackage{hyperref}

\begin{document}
\title{Light one-electron quasi-molecular ions within the finite-basis-set method for the two-center Dirac equation}
\author{D. Solovyev$^{1,2}$}
\email[E-mail:]{d.solovyev@spbu.ru}
\author{A. Anikin$^{1,3}$}
\author{A. Danilov$^{1}$}
\author{D. Glazov$^{1,4}$}
\author{A. Kotov$^1$}
\affiliation{ 
$^1$ Department of Physics, St. Petersburg State University, Petrodvorets, Oulianovskaya 1, 198504, St. Petersburg, Russia\\
$^2$ Petersburg Nuclear Physics Institute named by B.P. Konstantinov of National Research Centre 'Kurchatov Institut', St. Petersburg, Gatchina, 188300, Russia\\
$^3$ D.I. Mendeleev Institute for Metrology, St. Petersburg, 190005, Russia\\
$^4$ School of Physics and Engineering, ITMO University, Kronverkskiy pr. 49, 197101 St. Petersburg, Russia
}

\begin{abstract}
The electronic spectra of light one-electron quasi-molecular compounds H-H$^+$, He$^+$-He$^2+$ and He$^+$-H$^+$ are analyzed. To this end, the two-center Dirac equation is solved by the dual-kinetically balanced finite-basis-set method for axially symmetric systems termed as A-DKB. This method allows a complete relativistic consideration of these systems at fixed internuclear distances. A comparison of the obtained results with the nonrelativistic and relativistic calculations presented in the literature is performed. The advantages and disadvantages of the approach are discussed in details.
\end{abstract}

\maketitle

\section{Introduction}

The study of light molecular systems has a long history beginning with the advent of the quantum mechanics (QM) theory \cite{BG-QM}. Since the first works in this area in the early stages of QM, the accurate theoretical description of the electronic energy spectra in the field of two nuclei has become one of the fundamental problems, see \cite{Greiner-RQM} and references therein. Among others, the calculations of the equilibrium inter-nuclear distance for H$_2$ and H$_2^+$ molecules served to verify and develop quantum theory \cite{Pauling}. Theoretical efforts are also focused on a suitable description of the response of such systems to external electromagnetic fields \cite{Bell-1950}.

Consistently improving experimental techniques yield increase in the accuracy of measurements and pose a challenge for theory.
In turn, ever more precise theoretical methods of studying light molecular compounds expanded the field of research and brought exciting results.
The one-electron two-center problem directly concerns plasma physics where the proton scattering cross section on the hydrogen atom in the ground and excited states is of particular interest, see, e.g., \cite{Hemsworth_2009,Abdurakhmanov_2018,Leung2019} and references therein. 
Also, the two-center problem is relevant for a wide range of studies on the calculation of proton and anti-proton stopping power in hydrogen.
In particular, appropriate two-center methods were applied aimed at reaching agreement with experiments in different proton energy regions \cite{Bailey_2019}.

In addition to laboratory studies of quasi-molecular compounds, light binuclear systems are of particular importance in the context of the chemistry of the early Universe \cite{Dalgarno_2005}.
Astrophysical studies are regularly devoted to the calculation of molecular ions for a number of reasons, see \cite{Hirata-H2+}. Calculations of one-electron systems H$_2^+$ and He$^+-p$ are complicated by the fact that to implement the principle of local thermodynamic equilibrium all electronic states including the continuum should be considered as possible. Thus, for example, a quasi-molecular recombination mechanism based on an adiabatic multi-level representation has been suggested and applied to treat the formation of atomic hydrogen in the early Universe \cite{Kereselidze-2020}. 

Being problematic for an exact mathematical solution, the two-center problem has been reduced to the development of numerical methods. 
Pursuing a precise determination of electron energies in bound states, numerical calculations have yielded outstanding results, which allow us to theoretically prescribe binding energies with an accuracy of $10^{-14} - 10^{-13}$ for the low-lying states in the H$_2^+$ molecular ion \cite{Yang1991,Kullie2001,Ishikawa-2008,Mironova-2015}. 
Partly turning such highly accurate calculations into the state of the art, the theoretical description of light binuclear molecules \cite{Nogueira-2023} is merged with experimental data \cite{Alighanbari2020,Patra2020,Kortunov2021}, leading to an accurate determination of the fundamental physical constants, further theoretical development, improvement of experimental techniques, etc. 
Through continuous experimental and theoretical advances, light binuclear systems have become the object of competitive molecular clock construction where the relative systematic error can be reduced to $5 \times 10^{-17}$ at room temperature \cite{Schiller2014,Schiller_P}.

Thus, although widely available in modern literature, a detailed analysis of the electron energies in these compounds is still relevant in various fields. In this paper, we present the results of one-electron energy calculations of the (quasi-)molecular systems H$-p$, He$^+$-He$^{2+}$ and He$^+-p$ ($p$ stands for a proton) in the low-lying bound states. The computations were performed by the A-DKB method \cite{Johnson_Bspline,Shabaev_DKB,Rozenbaum_ADKB}, within the framework of which the high-lying energy states can also be obtained quite accurately. Without claiming the highest level of accuracy, the solution of the Dirac equation with a two-center potential within this approach can be considered as an alternative to other methods and, in particular, to one of the most accurate nonrelativistic calculations \cite{Bakalov2006,Korobov2014,Korobov2017}. 

The analysis and detailed comparison of the A-DKB results with the available literature is the main purpose of this study. 
The paper is organized as follows. 
In the next section, a brief description of the A-DKB method is given, followed by a discussion of the results for the one-electron molecular ion H$_2^+$. 
Numerical results for the quasi-molecular He$^{+}$-He$^{2+}$ and He$^+-p$ ions are analyzed in the sections~\ref{HeHe3+} and \ref{He+p}, respectively, and conclusions are given in the last section of the paper. We use the relativistic system of units $\hbar = c = m_e = 1$.

\section{The dual kinetic balance method for systems with axial symmetry (A-DKB)}

\begin{figure}[hbtp]
\centering
\includegraphics[width=0.8\columnwidth]{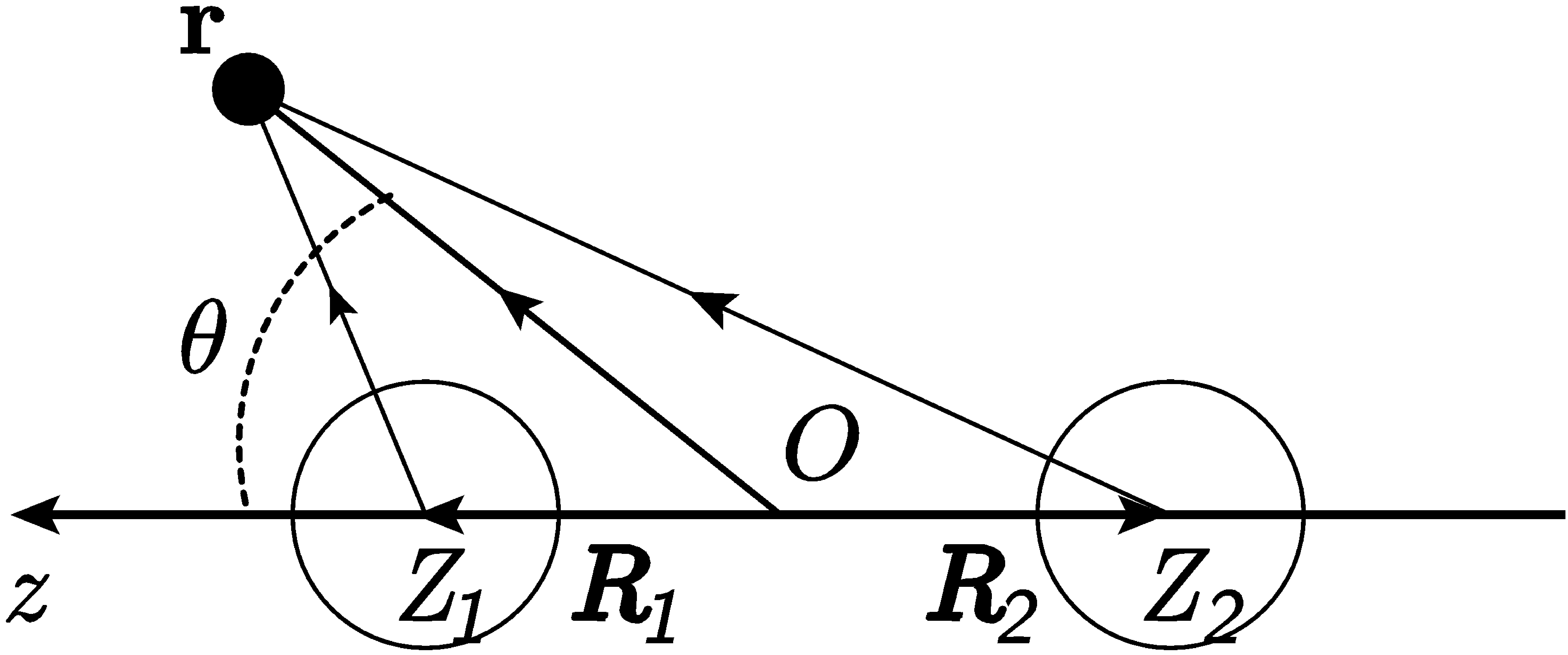}
\caption{The coordinate system for bi-nuclear one-electron compounds oriented along the $z$-axis. The bound electron is marked with a filled circle, the nuclei are marked with circles with the nucleus charges $Z_1$ and $Z_2$. The corresponding position vectors from the coordinate origin $O$ are indicated by $\boldsymbol{R}_1$ and $\boldsymbol{R}_2$. The position vector of the bound electron placed on one of the nuclei is given by $\boldsymbol{r}$ directed from $O$. Since axial symmetry, the azimuth angle $\theta$ is introduced. }
\label{fig1}
\end{figure}

In this paper, we consider simple systems consisting of two nuclei (heavy charged particles) and one electron. Representing a chemical compound that does not necessarily have an equilibrium inter-nuclear distance, such systems can form one-electron (quasi-)molecular ions. 
We restrict ourselves to the following compounds: the quasi-molecular ions H$-p$, He$^+$-He$^{2+}$ and He$^+-p$. To describe these systems, we also consider only the energy spectrum of the bound electron, which can effectively be obtained by solving the stationary Dirac equation in the Born-Oppenheimer approximation:
\begin{eqnarray}
\label{1}
\hat{H}_{\rm D}\psi_n= E_n\psi_n,
\end{eqnarray}
where
\begin{eqnarray}
\hat{H}_{\rm D} & = & \bm{\alpha}\cdot\mathbf{p}+\hat{V}(Z_1,Z_2,\bm{r})+\beta, \\
V_{\rm nucl}(Z_1,Z_2,\bm{r}) & = & V_{\rm nucl}(Z_1,\left|\bm{r}-\bm{R}_1\right|) +V_{\rm nucl}(Z_2,\left|\bm{r}-\bm{R}_2\right|)
\nonumber
\\
 & \equiv & -\frac{\alpha Z_1}{\left|\bm{r}-\bm{R}_1\right|}-\frac{\alpha Z_2}{\left|\bm{r}-\bm{R}_2\right|}.
\end{eqnarray}
Here, the nuclear binding potential operator $V_{\text{nucl}}$ and the electron wave functions, $\psi_n$, depend on the inter-nuclear distance $R$ as a parameter, $Z_1$ and $Z_2$ represent the charge number for the corresponding nucleus, the position vectors $\bm{r}$, $\bm{R}_1$ and $\bm{R}_2$ correspond to the electron and nuclei, respectively, and are determined according to the coordinate system scheme depicted in Fig.~{\ref{fig1}}.


Commonly the multipole partial-wave decomposition in the center-of-mass system is used,
 see, for example, \cite{Yang1991,Mironova-2015}. 
We use a method which permits to carry out calculations without such approximation and has shown its efficiency for completely relativistic calculations of heavy quasi-molecules \cite{Kotov2020,Kotov2021,Kotov2022}. 
The method is based on the decomposition of the required wave function over a finite basis set of B-splines \cite{Johnson_Bspline,Sapirstein_1996} with the imposed dual-kinetic-balance conditions \cite{Shabaev_DKB} generalized to the case of axially symmetric systems \cite{Rozenbaum_ADKB}.

In view of the axial symmetry, the required solution (bispinor) in the spherical coordinate system can be written in the form:
\begin{eqnarray}
\label{3}
\psi(r,\theta,\varphi) = \frac{1}{r}
\begin{pmatrix}
G_1(r,\theta) e^{\rm{i}(m_J-1/2)\varphi}
\\
G_2(r,\theta) e^{\rm{i}(m_J+1/2)\varphi}
\\
{\rm i}F_1(r,\theta) e^{\rm{i}(m_J-1/2)\varphi}
\\
{\rm i}F_2(r,\theta) e^{\rm{i}(m_J+1/2)\varphi}
\end{pmatrix},
\end{eqnarray}
where $m_J$ is the eigenvalue of the angular momentum operator $J_z$ (projection of the angular momentum $J$ onto the molecular axis $z$). The radial-azimuthal components $G_i(r,\theta)$ and $F_i(r,\theta)$ ($i=1,2$) through a finite basis set expansion are represented as
\begin{eqnarray}
\label{4}
\phi(r,\theta)=
\begin{pmatrix}
G_1(r,\theta)
\\
G_2(r,\theta)
\\
F_1(r,\theta)
\\
F_2(r,\theta)
\end{pmatrix}
\approx\sum\limits_{u=1}^4 
\sum\limits_{\substack{i_r=1 \\ i_\theta=1}}^{N_r, N_\theta} \Lambda C_{i_r i_\theta}^u B_{i_r}(r)Q_{i_\theta}(\theta)e_u,\qquad
\end{eqnarray}
where $B_{i_r}(r)$ is given by the B-splines, $Q_{i_\theta}$ are Legendre polynomials of the argument $\frac{2\theta}{\pi} - 1$, $e_u$ are the four-component basis unit vectors. 
According to \cite{Johnson_Bspline, Sapirstein_1996}, using a spectral approach based on B-splines shows that the basis (\ref{4}) leads to spurious states due to an unconstrained negative spectrum. To avoid this problem, a $\Lambda$ matrix is introduced which imposes certain, dual-kinetic-balance (DKB), relations between the upper and lower components of the Dirac bispinor derived from the nonrelativistic limit. 
This procedure reflects the DKB method, see details and discussion, as well as the explicit form of the matrix, in \cite{Shabaev_DKB,Rozenbaum_ADKB}.

Application of the DKB approach for axially symmetric systems is implied for the extended charge nucleus only. The point-like nucleus case can be accessed by the extrapolation of the extended-nucleus results in vanishing nuclear size. The expression above is given for arbitrary basis sets $\{B_{i_r}(r)\}_{i_r=1}^{N_r}$ and $\{Q_{i_\theta}(r)\}_{i_\theta=1}^{N_\theta}$. A particular choice of one-component basis functions corresponds to $B$-splines of the second order and Legendre polynomials as $Q_{i_\theta}(\theta) = P_{i_\theta-1}\left(\frac{2\theta}{\pi}-1\right)$, forming the set of one-component $\theta$-dependent basis functions of polynomial degrees $l=0\dots N_\theta-1$. By varying the values $N_r$, $N_\theta$, such a sample allows the convergence of numerical calculations to be monitored and, within a sufficiently small amount of compute time, guarantees about seven significant digits.

\section{${\rm H}-p\,$ quasi-molecular ion}
\label{H-p}
We begin with the well-studied system, which is a positive ion of a hydrogen molecule, H$_2^+$, see, for example, \cite{Carrington_1989}. Being entirely relativistic, the values obtained by the A-DKB method can be directly compared with the nonrelativistic results \cite{Korobov_2007}, with the aim of determining the relativistic corrections. The A-DKB values for the atomic spectrum of binding electron energies were attained on a grid with parameters $N_r=550$ and $N_\theta=80$. 

Numerical results for various inter-nuclear distances are collected in Table~\ref{tab:1} for the ground state in H$_2^+$ ion. In this table we also illustrate the nonrelativistic values borrowed from \cite{Korobov_2007} used to determine the relativistic energy shift, $\Delta E_{\rm rel}$. In order to compare the results, we also give the values of the relativistic leading-order correction \cite{Korobov_2006}, $\delta E_{\rm BP}$. The number of digits in the borrowed values is abbreviated for brevity.
\begin{table}[ht]
\begin{center}
\caption{The ground state binding energy of H$_2^+$ ion at the different inter-nuclear distances. All values are given in the atomic units, a.u.}
\label{tab:1}
\begin{tabular}{ c  c  c  c  c }
\hline
\hline
R & $E_{\rm A-DKB}$ & $E_{\rm nr}$ \cite{Korobov_2007} & $\Delta E_{\rm rel}$ & $\delta E_{\rm BP}$ \cite{Korobov_2006}\\
\hline
\noalign{\smallskip}
$0.1$ & $ -1.9783363 $ & $ -1.9782421$ 
 & $ -9.421\times 10^{-5}$ & $-9.420\times 10^{-5}$\\
\noalign{\smallskip}
$0.5$ & $ -1.7350283 $ & $ -1.7349880$
 & $ -4.026\times 10^{-5}$ & $-4.027\times 10^{-5}$\\
\noalign{\smallskip}
$1.0$ & $ -1.45180402 $ & $ -1.4517863$
 & $ -1.770\times 10^{-5}$ & $-1.769\times 10^{-5}$\\
\noalign{\smallskip}
$1.5$ & $ -1.2490002 $ & $ -1.2489899$
 & $ -1.249\times 10^{-5}$ & $--$\\
\noalign{\smallskip}
$2.0$ & $ -1.10264160 $ & $ -1.1026342$
 & $ -7.390\times 10^{-6}$ & $-7.366\times 10^{-6}$\\
\noalign{\smallskip}
$2.5$ & $ -0.99382959 $ & $ -0.9938235$
 & $ -6.081\times 10^{-6}$ & $--$\\
\noalign{\smallskip}
$3.0$ & $ -0.91090171 $ & $ -0.9108962$
 & $ -5.517\times 10^{-6}$ & $-5.482\times 10^{-6}$\\
\noalign{\smallskip}
$3.5$ & $ -0.84657515 $ & $ -0.8465698$
 & $ -5.331\times 10^{-6}$ & $--$\\
\noalign{\smallskip}
$4.0$ & $ -0.79609024 $ & $ -0.7960849$
 & $-5.358\times 10^{-6}$ & $-5.305\times 10^{-6}$\\
\noalign{\smallskip}
$5.0$ & $ -0.72442599 $ & $ -0.7244203$
 & $ -5.697\times 10^{-6}$ & $-5.625\times 10^{-6}$\\
\noalign{\smallskip}
$6.0$ & $ -0.67864182 $ & $ -0.6786357$
 & $ -6.102\times 10^{-6}$ & $-6.014\times 10^{-6}$\\
\noalign{\smallskip}
$7.0$ & $ -0.64845754 $ & $ -0.6484511$
 & $ -6.396\times 10^{-6}$ & $-6.306\times 10^{-6}$\\
\noalign{\smallskip}
$8.0$ & $ -0.6275769 $ & $ -0.6275704$
 & $ -6.551\times 10^{-6}$ & $-6.479\times 10^{-6}$\\
\noalign{\smallskip}
$9.0$ & $ -0.61231316 $ & $ -0.6123066$
 & $ -6.599\times 10^{-6}$ & $-6.569\times 10^{-6}$\\
\noalign{\smallskip}
$10.0$ & $ -0.60058529 $ & $ -0.6005787$
 & $ -6.569\times 10^{-6}$ & $-6.614\times 10^{-6}$\\
\hline
\hline
\end{tabular}
\end{center}
\end{table}

At first, we compare our result for the ground-state binding energy of the H$_2^+$ ion with the appropriate relativistic calculations at the inter-nuclear distance $R=2$ a.u. \cite{Ishikawa-2008,Mironova-2015}. 
The deviation constitutes $2.08\times 10^{-8}$ relative value, while the results of the mentioned works correlate with each other at the level of $10^{-12}$. Without the goal of high-precision calculation, we obtain that the chosen grid parameters give sufficient accuracy to extract the relativistic leading-order correction, which is about $\sim 10^{-5}$. 

Considering values of $\Delta E_{\rm rel}$ in Table~\ref{tab:1}, it is possible to notice the increasing difference between relativistic and purely non-relativistic results with decreasing $R$. 
Therefore the significance of the correction for the finite size of the nuclei can be assumed. We performed calculations for different values of the proton charge radius ($r_p=0.841$ fm and $r_p=0.877$ fm) combined with different models of the nucleus (Fermi and shell models) and found the corresponding dependence to be an order of magnitude smaller than the deviation of our results from \cite{Ishikawa-2008}, that is at $10^{-9}$ level. 
Such inaccuracy was obtained at all distances. 
In the A-DKB approach, we associate the deviation of the relativistic correction $\Delta E_{\rm rel}$ relative to $\delta E_{\rm BP}$ with the inaccuracy of our calculations. The absolute value of the relative deviation can be calculated using the last two columns of Table~\ref{tab:1} as $\left|\Delta E_{\rm rel}-\delta E_{\rm BP}\right|/\left|\Delta E_{\rm rel}\right|$. 
However, the accuracy can be improved by appropriate parameter enhancement or by matching a more suitable nonlinear grid. A direct increase in the basis $N_r$ and $N_\theta$ significantly extends the computation time, so we restrict ourselves to the current $10^{-4}-10^{-3}$ level.

As it follows from Table~\ref{tab:1}, the A-DKB method works well for closer nuclei and becomes worse with the growth of inter-nuclear distance. Despite the obvious improvement in numerical accuracy for small inter-nuclear distances, we assume further the same discrepancy for the results presented in Table~\ref{tab:2}, which collects data at a range of $[10,5000]$ fm.
\begin{table}[ht]
\begin{center}
\caption{The ground-state binding energy of H$_2^+$ ion as a function of inter-nuclear distance, $R$. The values of energies are given in atomic units, while the inter-nuclear distances are given in fermi, fm.}
\label{tab:2}
\begin{tabular}{ c  c  c  c }
\hline
\hline
R, fm & $E_{\rm A-DKB}$, a.u. & $R$, fm & $E_{\rm A-DKB}$, a.u. \\
\hline
\noalign{\smallskip}
$10$ & $ -2.000106416 $ & $2000$ & $ -1.996575311 $ \\
\noalign{\smallskip}
$50$ & $ -2.000104131 $ & $2500$ & $ -1.994694718 $ \\
\noalign{\smallskip}
$100$ & $ -2.000097011 $ & $3000$ & $ -1.992462892 $\\
\noalign{\smallskip}
$300$ & $ -2.000021675 $ & $3500$ & $ -1.989902025 $ \\
\noalign{\smallskip}
$500$ & $ -1.999872666 $ & $4000$ & $ -1.987033284 $ \\
\noalign{\smallskip}
$1000$ & $ -1.999188963 $ & $4500$ & $ -1.983876820 $ \\
\noalign{\smallskip}
$1500$ & $ -1.998081459 $ & $5000$ & $ -1.980451775 $ \\

\hline
\hline
\end{tabular}
\end{center}
\end{table}

\begin{figure}[hbtp]
\centering
\includegraphics[width=0.8\columnwidth]{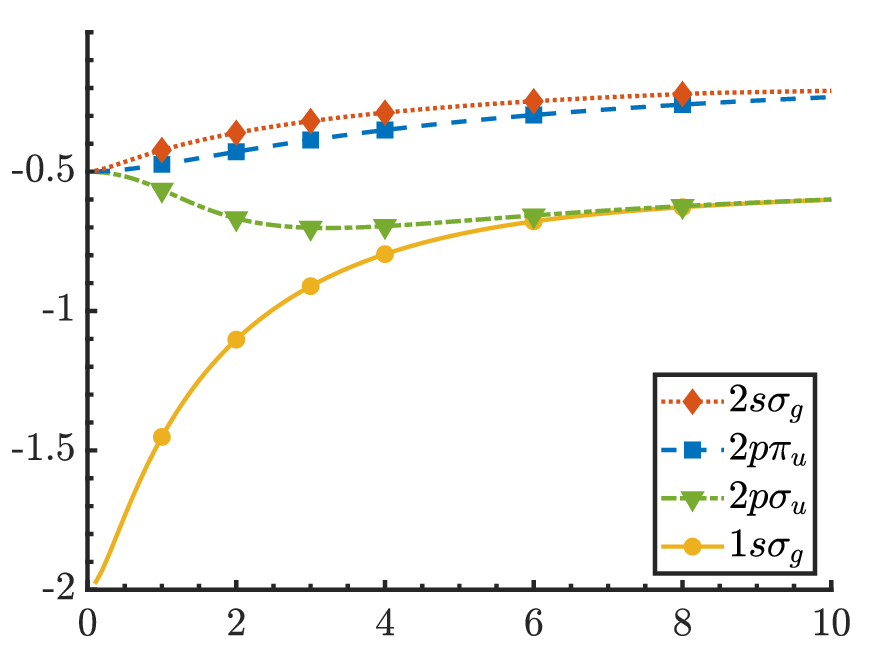}
\caption{The adiabatic potential curves of the first four lower states for the $\mathrm{H} - p$ quasi-molecular ion in dependence on inter-nuclear distance. 
All values are given in atomic units.}
\label{fig3}
\end{figure}

In addition to the numerical results for the ground state in H$^+_2$ molecular ion, we present the adiabatic potential curves for the first four states, see Fig.~\ref{fig3}.

In particular, the graph clearly shows the splitting of atomic energy levels as the nuclei approach \cite{QC-Eyring}. 
As expected, the diving or pulling out of the energy levels is related to their symmetry. 
The binding energies goes to the neutral hydrogen energy states with the increasing inter-nuclear distance and become equal to atomic energies with $Z+1$ nuclear charge at $R=0$, which we denote here as $^2$He$^+$ ion (two protons being nucleons). 
According to Fig.~\ref{fig3} the lowest molecular state $1\sigma_g$ goes to the ground energy level $1s_{1/2}$ in the atomic ion $^2$He$^+$ at $R\rightarrow 0$. The three intermediate energy levels of the molecule at the middle distances in Fig.~\ref{fig3} degenerate at $R\rightarrow 0$, giving an energy state with the principal quantum number $n=2$ in the $^2$He$^+$ atom in the non-relativistic limit. 

\begin{figure}[hbtp]
\centering
\includegraphics[width=0.8\columnwidth]{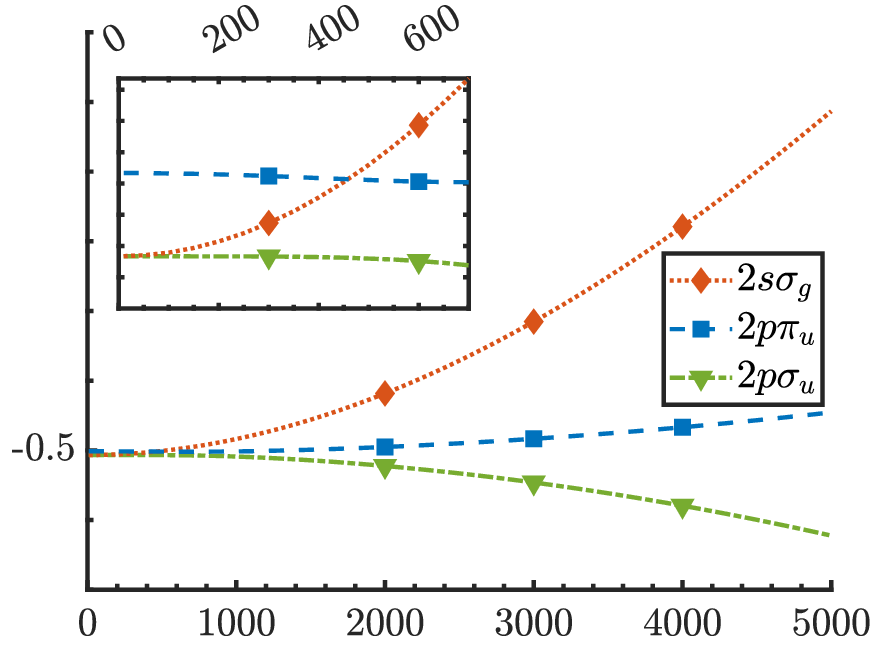}
\caption{Adiabatic potential curves of the first excited state with $n=2$ of the quasi-molecular ion $\mathrm{H} - p$ for inter-nuclear distances in the range $[10,5000]$ fm. All values are in atomic units.}
\label{fig4}
\end{figure}

It also follows from Fig.~\ref{fig3} that the degenerate level of the hydrogen atom with $n=2$ passes into a similar level in a one-electron $^2$He$^+$ ion. In the non-relativistic limit these levels (at zero and at infinity) are degenerate. In the relativistic picture there is a fine splitting corresponding to different total angular momentum of the atomic electron. At the same time, states with the same momentum (such as $2s_{1/2}$ and $2p_{1/2}$) remain degenerate without taking into account QED effects (Lamb shift in particular). Our calculations have revealed that at distances on the order of $500$ fm the level crossing occurs, bringing the hydrogen level $2p_{3/2}$ into the degenerate $2s_{1/2}/2p_{1/2}$. The scheme of levels with $n=2$ calculated using Dirac theory within the A-DKB method in the range $[10,5000]$ fm is demonstrated in Fig.~\ref{fig4}, where the intersection of the levels is explicitly shown in the inset.
In particular, combining the graphs in Figs.~\ref{fig3} and \ref{fig4} allows us to trace clearly the splitting of the ground and first excited states, as well as the subsequent evolution of quasi-molecular terms into the ground and first excited degenerate states in the one-electron ion.

\section{${\rm He}^+-{\rm He}^{2+}$ quasi-molecular ion}
\label{HeHe3+}
\begin{figure}[hbtp]
\centering
\includegraphics[width=0.8\columnwidth]{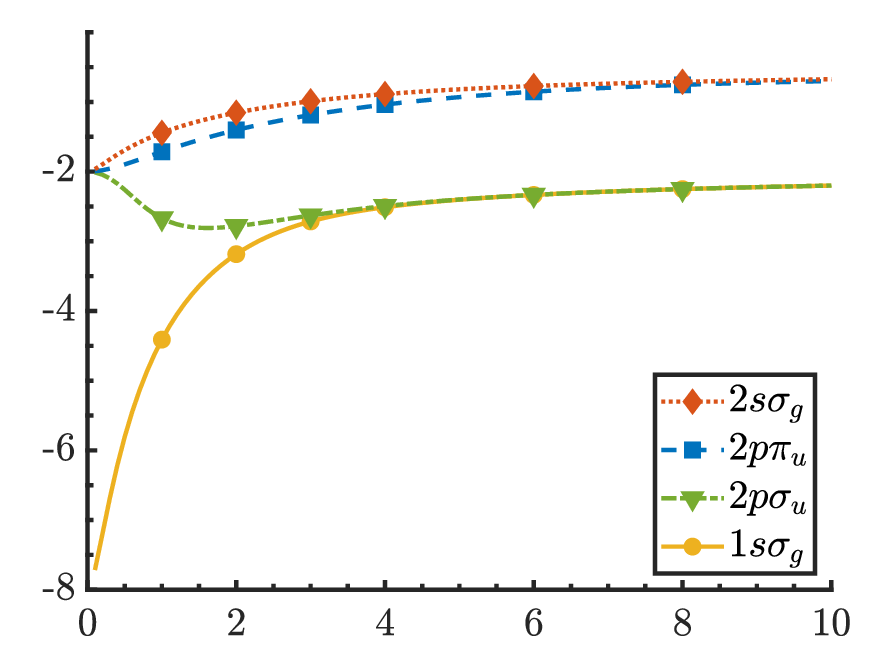}
\caption{Adiabatic potential curves of the first four lower states for the quasi-molecular ion $\mathrm{He}_2^{3+}$ depending on the inter-nuclear distance. All values are in a.u.}
\label{fig5}
\end{figure}

As a continuation of our study, we consider the homonuclear quasi-molecular ion corresponding to the one-electron compound ${\rm He}^+-{\rm He}^{2+}$, i.e., the atomic He$^+$ ion and the bare helium nucleus. The resulting energies as a function of the inter-nuclear distance for the ground state are shown in Table~\ref{tab:3}. We estimate the uncertainty of our numerical calculations under the A-DKB approach for this compound, obtained with the same grid parameters as before, i.e., at the level of $10^{-8}$. In Fig.~\ref{fig5} we also illustrate the behavior of the first four energy levels of the bound electron in this compound as a function of the inter-nuclear distance.
\begin{table}[hbtp]
\begin{center}
\caption{The ground state energy of He$_2^{3+}$ as a function of the inter-nuclear distance, $R$. All values are given in atomic units.}
\label{tab:3}
\begin{tabular}{ c  c  c  c }
\hline
\hline
R, a.u. & $E_{\rm A-DKB}$, a.u. & $R$, a.u. & $E_{\rm A-DKB}$, a.u. \\
\hline
\noalign{\smallskip}
$0.1$ & $ -7.715707556 $ & $4.0$ & $ -2.510384560 $ \\
\noalign{\smallskip}
$0.5$ & $ -5.807428344 $ & $5.0$ & $ -2.402419325 $ \\
\noalign{\smallskip}
$1.0$ & $ -4.410654719 $ & $6.0$ & $ -2.334110182 $\\
\noalign{\smallskip}
$1.5$ & $ -3.643672476 $ & $7.0$ & $ -2.286091796 $ \\
\noalign{\smallskip}
$2.0$ & $ -3.184424360 $ & $8.0$ & $ -2.250244278 $ \\
\noalign{\smallskip}
$2.5$ & $ -2.897771046 $ & $9.0$ & $ -2.222404767 $ \\
\noalign{\smallskip}
$3.0$ & $ -2.714638840 $ & $10.0$ & $ -2.200146300 $ \\

\hline
\hline
\end{tabular}
\end{center}
\end{table}

Numerical results for electron energies in the range of inter-nuclear distances from 10 to 5000 fm are given in Table~\ref{tab:4} for the ground state of the $\mathrm{He}_2^{3+ }$ molecular ion.
\begin{table}[ht]
\begin{center}
\caption{The ground state energy (in a.u.) of $\mathrm{He}_2^{3+}$ ion depending on the inter-nuclear distance, $R$ (in fm). 
}
\label{tab:4}
\begin{tabular}{ c  c  c  c }
\hline
\hline
R, fm & $E_{\rm A-DKB}$, a.u. & $R$, fm & $E_{\rm A-DKB}$, a.u. \\
\hline
\noalign{\smallskip}
$10$ & $ -8.001703064 $ & $2000$ & $ -7.949313335 $ \\
\noalign{\smallskip}
$50$ & $ -8.001666435 $ & $2500$ & $ -7.922949204 $ \\
\noalign{\smallskip}
$100$ & $ -8.001552614 $ & $3000$ & $ -7.892572931 $\\
\noalign{\smallskip}
$300$ & $ -8.000358658 $ & $3500$ & $ -7.858721599 $ \\
\noalign{\smallskip}
$500$ & $ -7.998024224 $ & $4000$ & $ -7.821875165 $ \\
\noalign{\smallskip}
$1000$ & $ -7.987546923 $ & $4500$ & $ -7.782461130 $ \\
\noalign{\smallskip}
$1500$ & $ -7.971066910 $ & $5000$ & $ -7.740859283 $ \\

\hline
\hline
\end{tabular}
\end{center}
\end{table}
The scheme of levels in this compound repeats the diproton structure but is confined to the region of small inter-nuclear distances because the nuclear charges $Z_1$ and $Z_2$ are two times greater than those of the molecule H$_2^+$.

The results collected in Table~\ref{tab:3} can be tested on the purely non-relativistic scaling factor of the electron energies, which is determined by the total charge of the nuclei $Z$ \cite{Pauling1933}. Using the values from Table~\ref{tab:1}, one can find the deviation from this linear scaling. The difference in behavior with respect to the non-relativistic scaling is explained by relativistic corrections, which are taken into account in our calculations. Thus, a detailed comparison of this compound and the H$_2^+$ ion can be used to test relativistic theory.

\section{${\rm He}^+-p$ quasi-molecular ion}
\label{He+p}
Within the framework of A-DKB method verification we consider also the simplest heteronuclear molecular ion ${\rm He}^+-p$. In contrast to the compound $\mathrm{He}_2^{3+}$, whose discussion in the literature is rather sparse (since it does not produce a bound compound), this system has been studied for years, see, for example, \cite{Janev_1997,Baxter2016}. To perform A-DKB calculations of electron binding energies for the ${\rm He}^+-p$ compound, we chose grid parameters $N_r=550$ and $N_\theta=80$, which guaranteed $10^{- 8}$ accuracy in the diproton ion. The numerical results are given in Table~\ref{tab:5} for various inter-nuclear distances.
\begin{table}[ht]
\begin{center}
\caption{The ground state energy of the $\mathrm{He}^{+} - p$ molecular ion at various inter-nuclear distances. All values are given in atomic units.}
\label{tab:5}
\begin{tabular}{c c c c }
\hline 
\hline
$R$, a.u. & $E_{\rm A-DKB}$, a.u. & $R$, a.u. & $E_{\rm A-DKB}$, a.u. \\ 
\hline
        
0.1  & $ -4.411760188 $  & 4.0  & $ -2.250710877 $ \\ 
0.5  & $ -3.665712855 $  & 5.0  & $ -2.200341640 $ \\ 
1.0  & $ -3.033459156 $  & 6.0  & $ -2.166881968 $ \\ 
1.5  & $ -2.695558820 $  & 7.0  & $ -2.143017926 $ \\ 
2.0  & $ -2.512296432 $  & 8.0  & $ -2.125132408 $ \\ 
2.5  & $ -2.404992440 $  & 9.0  & $ -2.111225690 $ \\ 
3.0  & $ -2.335596530 $  & 10.0 & $ -2.100100978 $ \\ 
\hline
\hline
\end{tabular}
\end{center}
\end{table}

\begin{figure}[hbtp]
\centering
\includegraphics[width=0.8\columnwidth]{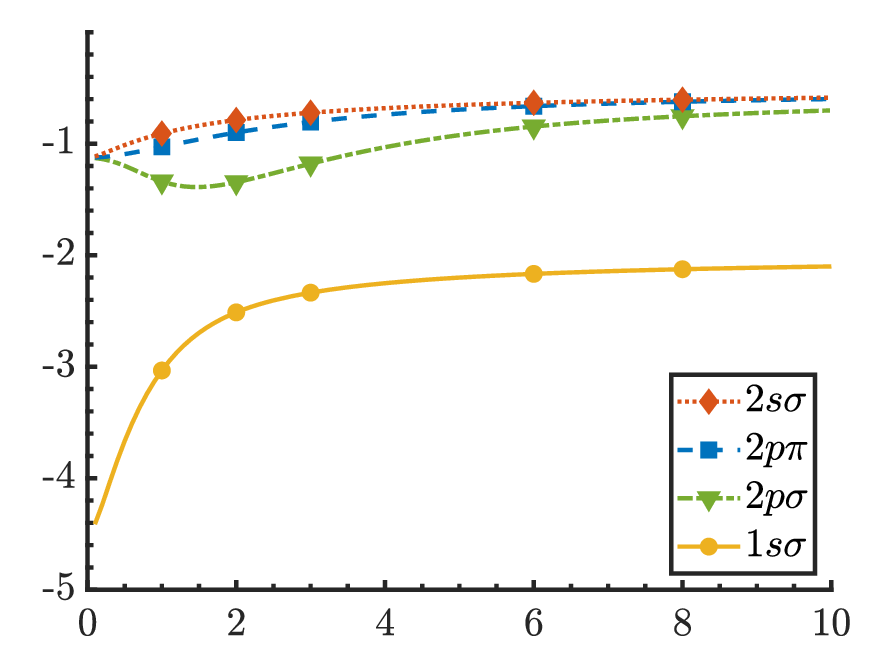}
\caption{Adiabatic potential curves of the first four lower states for the quasi-molecular ion $\mathrm{He}^{+} - p$ depending on the inter-nuclear distance. All values are in a.u.}
\label{fig6}
\end{figure}

Adiabatic curves for low binding energies in the quasi-molecule $\mathrm{He}^{+} - p$ are illustrated in Fig.~\ref{fig6} repeating the behaviour presented in \cite{Janev_1997}.
As can be seen from comparing the plots in Figs.~\ref{fig4}, \ref{fig5}, the adiabatic potential curves are essentially the same. The only difference is in the region of inter-nuclear distances where the levels of molecular ions split and, conversely, degenerate. These distances are determined by a parameter related to the charges of the nuclei, so Fig.~\ref{fig6} is an intermediate link between Fig.~\ref{fig3} and Fig.~\ref{fig5}.

Finally, below we give the values of the ground state energy in $\mathrm{He}^{+} - p$ molecular ion at inter-nuclear distances at a range $[10,5000]$ fm.
\begin{table}[ht]
\begin{center}
\caption{The ground state energy (in a.u.) of the $\mathrm{He}^{+} - p$ molecular ion at small inter-nuclear distances $R$ (in fm).
}
\label{tab:6}
\begin{tabular}{c c c c }
\hline 
\hline
$R$, fm & $E_{\rm A-DKB}$, a.u. & $R$, fm & $E_{\rm A-DKB}$, a.u. \\ 
\hline
        
10  & $ -4.500538828 $  & 2000  & $ -4.485233634 $ \\ 
50  & $ -4.500528540 $  & 2500  & $ -4.477305988 $ \\ 
100  & $ -4.500496521 $  & 3000  & $ -4.468034351 $ \\ 
300  & $ -4.500159173 $  & 3500  & $ -4.457549106 $ \\ 
500  & $ -4.499495719 $  & 4000  & $ -4.445970839 $ \\ 
1000  & $ -4.496484507 $  & 4500  & $ -4.433410776 $ \\ 
1500  & $ -4.491676759 $  & 5000 & $ -4.419971290 $ \\ 
\hline
\hline
\end{tabular}
\end{center}
\end{table}

\section{Conclusions}

The main goal of the present study consists in verification of the A-DKB method \cite{Rozenbaum_ADKB} on the light molecular compounds. The validity of the A-DKB approach has been tested on various one-electron ions: $\mathrm{H}_2^{+}$, $\mathrm{He}_2^{3+}$ and ${\rm He}^ +-p$, that is, including homo- and heteronuclear molecules. Although the method used here was adapted to study quasi-molecules with heavy nuclei \cite{Kotov2020,Kotov2021,Kotov2022}, the obtained results show sufficient accuracy with the data presented in the literature. We estimated the accuracy of our values at the level of $10^{-8}$, which was obtained by comparison with numerical results obtained within the non-relativistic approach \cite{Korobov_2006,Korobov_2007}. Since our calculations are carried out within a fully relativistic Dirac picture, we found a relativistic leading-order correction with an accuracy of about $10^{-3}$ relative magnitude for the $\mathrm{H}_2^{+}$ compound. Thus, the molecular ion  $\mathrm{H}_2^{+}$ was used to calibrate the grid parameters utilized further for the calculations of $\mathrm{He}_2^{3+}$ and ${\rm He}^ +-p$ molecular systems.

Within our study, we have constructed the adiabatic potential curves for all mentioned molecules for the first four energy terms and presented the numerical data for the ground states as a function of inter-nuclear distance, $R$. Although $\mathrm{H}_2^{+}$ is widely discussed in the literature and highly accurate calculations for this system yield about twenty significant digits \cite{Ishikawa-2008,Mironova-2015} in the relativistic approach, on its grounds we presented data for ions $\mathrm{He}_2^{3+}$ and ${\rm He}^ +-p$, that are barely found in the literature. In addition, our calculations also make it possible to take into account highly excited states, which can be used in various effects described in the perturbation theory of second (and higher) order.

The A-DKB method shows good agreement with other approaches at large inter-nuclear distances (for $R=10$ a.u. the accuracy is on the order of $10^{-8} - 10^{-7}$). But for small inter-nuclear distances the accuracy of our calculations is higher. In this regard, for all of the compounds considered in this paper, we have given values of the ground state energy as a function of the inter-nuclear distance in the range from 10 to 5000 fm. According to Fig.~\ref{fig4}, there is an intersection of the first two excited states at small distances. For the diproton compound, this intersection occurs at $R\approx 500$ fm, and is valid for the He$_2^{3+}$ molecular ion. This picture holds for any homonuclear molecules, resulting in crossing states at small distances, weakly depending on the total nuclear charge $Z_1 +Z_2$ (for $Z_1,Z_2=50$ this distance is about $450$ fm). 
From Figs.~\ref{fig3}, \ref{fig5}, and \ref{fig6} it is clearly seen that the molecular terms degenerate into nonrelativistic energies at $R\rightarrow 0$.

Finally, the results for the considered homonuclear quasi-molecules can be used to study relativistic effects by the comparison of our calculations with the purely nonrelativistic scaling factor $Z=Z_1+Z_2$. The linear dependence preserves only for small distances $\leq 1000$ fm, see Tables~\ref{tab:2}, \ref{tab:4} and \ref{tab:6}. In addition to laboratory studies involving, for example, charge transfer phenomena \cite{Minami_2007}, collisions of He atoms \cite{Baxter2016,Faulkner_2019} or atomic data formation \cite{angeli2013table}, we hope that the obtained here results will also be in demand in astrophysical research, see, for example, \cite{Furlanetto2007,Ravi2021}.

\textit{Acknowledgements.} This work was supported by RSF grant N\textsuperscript{\underline{o}} 23-22-00250.
\bibliographystyle{apsrev4-1}
\bibliography{mybibfile}  

\begin{thebibliography}{43}%
\makeatletter
\providecommand \@ifxundefined [1]{%
 \@ifx{#1\undefined}
}%
\providecommand \@ifnum [1]{%
 \ifnum #1\expandafter \@firstoftwo
 \else \expandafter \@secondoftwo
 \fi
}%
\providecommand \@ifx [1]{%
 \ifx #1\expandafter \@firstoftwo
 \else \expandafter \@secondoftwo
 \fi
}%
\providecommand \natexlab [1]{#1}%
\providecommand \enquote  [1]{``#1''}%
\providecommand \bibnamefont  [1]{#1}%
\providecommand \bibfnamefont [1]{#1}%
\providecommand \citenamefont [1]{#1}%
\providecommand \href@noop [0]{\@secondoftwo}%
\providecommand \href [0]{\begingroup \@sanitize@url \@href}%
\providecommand \@href[1]{\@@startlink{#1}\@@href}%
\providecommand \@@href[1]{\endgroup#1\@@endlink}%
\providecommand \@sanitize@url [0]{\catcode `\\12\catcode `\$12\catcode
  `\&12\catcode `\#12\catcode `\^12\catcode `\_12\catcode `\%12\relax}%
\providecommand \@@startlink[1]{}%
\providecommand \@@endlink[0]{}%
\providecommand \url  [0]{\begingroup\@sanitize@url \@url }%
\providecommand \@url [1]{\endgroup\@href {#1}{\urlprefix }}%
\providecommand \urlprefix  [0]{URL }%
\providecommand \Eprint [0]{\href }%
\providecommand \doibase [0]{http://dx.doi.org/}%
\providecommand \selectlanguage [0]{\@gobble}%
\providecommand \bibinfo  [0]{\@secondoftwo}%
\providecommand \bibfield  [0]{\@secondoftwo}%
\providecommand \translation [1]{[#1]}%
\providecommand \BibitemOpen [0]{}%
\providecommand \bibitemStop [0]{}%
\providecommand \bibitemNoStop [0]{.\EOS\space}%
\providecommand \EOS [0]{\spacefactor3000\relax}%
\providecommand \BibitemShut  [1]{\csname bibitem#1\endcsname}%
\let\auto@bib@innerbib\@empty
\bibitem [{\citenamefont {Bromley}\ and\ \citenamefont
  {Greiner}(2000)}]{BG-QM}%
  \BibitemOpen
  \bibfield  {author} {\bibinfo {author} {\bibfnamefont {D.}~\bibnamefont
  {Bromley}}\ and\ \bibinfo {author} {\bibfnamefont {W.}~\bibnamefont
  {Greiner}},\ }\href@noop {} {\emph {\bibinfo {title} {Quantum Mechanics: An
  Introduction}}},\ Physics and Astronomy\ (\bibinfo  {publisher} {Springer
  Berlin Heidelberg},\ \bibinfo {year} {2000})\BibitemShut {NoStop}%
\bibitem [{\citenamefont {Greiner}(2000)}]{Greiner-RQM}%
  \BibitemOpen
  \bibfield  {author} {\bibinfo {author} {\bibfnamefont {W.}~\bibnamefont
  {Greiner}},\ }\href@noop {} {\emph {\bibinfo {title} {Relativistic Quantum
  Mechanics. Wave Equations}}}\ (\bibinfo  {publisher} {Springer},\ \bibinfo
  {year} {2000})\BibitemShut {NoStop}%
\bibitem [{\citenamefont {Pauling}(1928)}]{Pauling}%
  \BibitemOpen
  \bibfield  {author} {\bibinfo {author} {\bibfnamefont {L.}~\bibnamefont
  {Pauling}},\ }\href {\doibase https://doi.org/10.1021/cr60018a003} {\bibfield
   {journal} {\bibinfo  {journal} {Chemical Reviews}\ }\textbf {\bibinfo
  {volume} {5}},\ \bibinfo {pages} {173} (\bibinfo {year} {1928})}\BibitemShut
  {NoStop}%
\bibitem [{\citenamefont {Bell}\ and\ \citenamefont {Long}(1950)}]{Bell-1950}%
  \BibitemOpen
  \bibfield  {author} {\bibinfo {author} {\bibfnamefont {R.~P.}\ \bibnamefont
  {Bell}}\ and\ \bibinfo {author} {\bibfnamefont {D.~A.}\ \bibnamefont
  {Long}},\ }\href {\doibase 10.1098/rspa.1950.0144} {\bibfield  {journal}
  {\bibinfo  {journal} {Proceedings of the Royal Society of London. Series A.
  Mathematical and Physical Sciences}\ }\textbf {\bibinfo {volume} {203}},\
  \bibinfo {pages} {364} (\bibinfo {year} {1950})}\BibitemShut {NoStop}%
\bibitem [{\citenamefont {Hemsworth}\ \emph {et~al.}(2009)\citenamefont
  {Hemsworth}, \citenamefont {Decamps}, \citenamefont {Graceffa}, \citenamefont
  {Schunke}, \citenamefont {Tanaka}, \citenamefont {Dremel}, \citenamefont
  {Tanga}, \citenamefont {Esch}, \citenamefont {Geli}, \citenamefont {Milnes},
  \citenamefont {Inoue}, \citenamefont {Marcuzzi}, \citenamefont {Sonato},\
  and\ \citenamefont {Zaccaria}}]{Hemsworth_2009}%
  \BibitemOpen
  \bibfield  {author} {\bibinfo {author} {\bibfnamefont {R.}~\bibnamefont
  {Hemsworth}}, \bibinfo {author} {\bibfnamefont {H.}~\bibnamefont {Decamps}},
  \bibinfo {author} {\bibfnamefont {J.}~\bibnamefont {Graceffa}}, \bibinfo
  {author} {\bibfnamefont {B.}~\bibnamefont {Schunke}}, \bibinfo {author}
  {\bibfnamefont {M.}~\bibnamefont {Tanaka}}, \bibinfo {author} {\bibfnamefont
  {M.}~\bibnamefont {Dremel}}, \bibinfo {author} {\bibfnamefont
  {A.}~\bibnamefont {Tanga}}, \bibinfo {author} {\bibfnamefont {H.~D.}\
  \bibnamefont {Esch}}, \bibinfo {author} {\bibfnamefont {F.}~\bibnamefont
  {Geli}}, \bibinfo {author} {\bibfnamefont {J.}~\bibnamefont {Milnes}},
  \bibinfo {author} {\bibfnamefont {T.}~\bibnamefont {Inoue}}, \bibinfo
  {author} {\bibfnamefont {D.}~\bibnamefont {Marcuzzi}}, \bibinfo {author}
  {\bibfnamefont {P.}~\bibnamefont {Sonato}}, \ and\ \bibinfo {author}
  {\bibfnamefont {P.}~\bibnamefont {Zaccaria}},\ }\href {\doibase
  10.1088/0029-5515/49/4/045006} {\bibfield  {journal} {\bibinfo  {journal}
  {Nuclear Fusion}\ }\textbf {\bibinfo {volume} {49}},\ \bibinfo {pages}
  {045006} (\bibinfo {year} {2009})}\BibitemShut {NoStop}%
\bibitem [{\citenamefont {Abdurakhmanov}\ \emph {et~al.}(2018)\citenamefont
  {Abdurakhmanov}, \citenamefont {Alladustov}, \citenamefont {Bailey},
  \citenamefont {Kadyrov},\ and\ \citenamefont {Bray}}]{Abdurakhmanov_2018}%
  \BibitemOpen
  \bibfield  {author} {\bibinfo {author} {\bibfnamefont {I.~B.}\ \bibnamefont
  {Abdurakhmanov}}, \bibinfo {author} {\bibfnamefont {S.~U.}\ \bibnamefont
  {Alladustov}}, \bibinfo {author} {\bibfnamefont {J.~J.}\ \bibnamefont
  {Bailey}}, \bibinfo {author} {\bibfnamefont {A.~S.}\ \bibnamefont {Kadyrov}},
  \ and\ \bibinfo {author} {\bibfnamefont {I.}~\bibnamefont {Bray}},\ }\href
  {\doibase 10.1088/1361-6587/aad436} {\bibfield  {journal} {\bibinfo
  {journal} {Plasma Physics and Controlled Fusion}\ }\textbf {\bibinfo {volume}
  {60}},\ \bibinfo {pages} {095009} (\bibinfo {year} {2018})}\BibitemShut
  {NoStop}%
\bibitem [{\citenamefont {Leung}\ and\ \citenamefont
  {Kirchner}(2019)}]{Leung2019}%
  \BibitemOpen
  \bibfield  {author} {\bibinfo {author} {\bibfnamefont {A.~C.~K.}\
  \bibnamefont {Leung}}\ and\ \bibinfo {author} {\bibfnamefont
  {T.}~\bibnamefont {Kirchner}},\ }\href {\doibase
  https://doi.org/10.1140/epjd/e2019-100380-x} {\bibfield  {journal} {\bibinfo
  {journal} {The European Physical Journal D}\ }\textbf {\bibinfo {volume}
  {73}},\ \bibinfo {pages} {246} (\bibinfo {year} {2019})}\BibitemShut
  {NoStop}%
\bibitem [{\citenamefont {Bailey}\ \emph {et~al.}(2019)\citenamefont {Bailey},
  \citenamefont {Abdurakhmanov}, \citenamefont {Kadyrov}, \citenamefont
  {Bray},\ and\ \citenamefont {Mukhamedzhanov}}]{Bailey_2019}%
  \BibitemOpen
  \bibfield  {author} {\bibinfo {author} {\bibfnamefont {J.~J.}\ \bibnamefont
  {Bailey}}, \bibinfo {author} {\bibfnamefont {I.~B.}\ \bibnamefont
  {Abdurakhmanov}}, \bibinfo {author} {\bibfnamefont {A.~S.}\ \bibnamefont
  {Kadyrov}}, \bibinfo {author} {\bibfnamefont {I.}~\bibnamefont {Bray}}, \
  and\ \bibinfo {author} {\bibfnamefont {A.~M.}\ \bibnamefont
  {Mukhamedzhanov}},\ }\href {\doibase 10.1103/PhysRevA.99.042701} {\bibfield
  {journal} {\bibinfo  {journal} {Phys. Rev. A}\ }\textbf {\bibinfo {volume}
  {99}},\ \bibinfo {pages} {042701} (\bibinfo {year} {2019})}\BibitemShut
  {NoStop}%
\bibitem [{\citenamefont {Dalgarno}(2005)}]{Dalgarno_2005}%
  \BibitemOpen
  \bibfield  {author} {\bibinfo {author} {\bibfnamefont {A.}~\bibnamefont
  {Dalgarno}},\ }\href {\doibase 10.1088/1742-6596/4/1/002} {\bibfield
  {journal} {\bibinfo  {journal} {Journal of Physics: Conference Series}\
  }\textbf {\bibinfo {volume} {4}},\ \bibinfo {pages} {10} (\bibinfo {year}
  {2005})}\BibitemShut {NoStop}%
\bibitem [{\citenamefont {Hirata}\ and\ \citenamefont
  {Padmanabhan}(2006)}]{Hirata-H2+}%
  \BibitemOpen
  \bibfield  {author} {\bibinfo {author} {\bibfnamefont {C.~M.}\ \bibnamefont
  {Hirata}}\ and\ \bibinfo {author} {\bibfnamefont {N.}~\bibnamefont
  {Padmanabhan}},\ }\href {\doibase 10.1111/j.1365-2966.2006.10924.x}
  {\bibfield  {journal} {\bibinfo  {journal} {Monthly Notices of the Royal
  Astronomical Society}\ }\textbf {\bibinfo {volume} {372}},\ \bibinfo {pages}
  {1175} (\bibinfo {year} {2006})}\BibitemShut {NoStop}%
\bibitem [{\citenamefont {Kereselidze}\ \emph {et~al.}(2020)\citenamefont
  {Kereselidze}, \citenamefont {Noselidze},\ and\ \citenamefont
  {Ogilvie}}]{Kereselidze-2020}%
  \BibitemOpen
  \bibfield  {author} {\bibinfo {author} {\bibfnamefont {T.}~\bibnamefont
  {Kereselidze}}, \bibinfo {author} {\bibfnamefont {I.}~\bibnamefont
  {Noselidze}}, \ and\ \bibinfo {author} {\bibfnamefont {J.~F.}\ \bibnamefont
  {Ogilvie}},\ }\href {\doibase 10.1093/mnras/staa3622} {\bibfield  {journal}
  {\bibinfo  {journal} {Monthly Notices of the Royal Astronomical Society}\
  }\textbf {\bibinfo {volume} {501}},\ \bibinfo {pages} {1160} (\bibinfo {year}
  {2020})}\BibitemShut {NoStop}%
\bibitem [{\citenamefont {Yang}\ \emph {et~al.}(1991)\citenamefont {Yang},
  \citenamefont {Heinemann},\ and\ \citenamefont {Kolb}}]{Yang1991}%
  \BibitemOpen
  \bibfield  {author} {\bibinfo {author} {\bibfnamefont {L.}~\bibnamefont
  {Yang}}, \bibinfo {author} {\bibfnamefont {D.}~\bibnamefont {Heinemann}}, \
  and\ \bibinfo {author} {\bibfnamefont {D.}~\bibnamefont {Kolb}},\ }\href
  {\doibase 10.1016/0009-2614(91)87058-J} {\bibfield  {journal} {\bibinfo
  {journal} {Chemical Physics Letters}\ }\textbf {\bibinfo {volume} {178}},\
  \bibinfo {pages} {213} (\bibinfo {year} {1991})}\BibitemShut {NoStop}%
\bibitem [{\citenamefont {Kullie}\ and\ \citenamefont
  {Kolb}(2001)}]{Kullie2001}%
  \BibitemOpen
  \bibfield  {author} {\bibinfo {author} {\bibfnamefont {O.}~\bibnamefont
  {Kullie}}\ and\ \bibinfo {author} {\bibfnamefont {D.}~\bibnamefont {Kolb}},\
  }\href {\doibase https://doi.org/10.1007/s100530170019} {\bibfield  {journal}
  {\bibinfo  {journal} {The European Physical Journal D - Atomic, Molecular,
  Optical and Plasma Physics}\ }\textbf {\bibinfo {volume} {17}},\ \bibinfo
  {pages} {167} (\bibinfo {year} {2001})}\BibitemShut {NoStop}%
\bibitem [{\citenamefont {Ishikawa}\ \emph {et~al.}(2008)\citenamefont
  {Ishikawa}, \citenamefont {Nakashima},\ and\ \citenamefont
  {Nakatsuji}}]{Ishikawa-2008}%
  \BibitemOpen
  \bibfield  {author} {\bibinfo {author} {\bibfnamefont {A.}~\bibnamefont
  {Ishikawa}}, \bibinfo {author} {\bibfnamefont {H.}~\bibnamefont {Nakashima}},
  \ and\ \bibinfo {author} {\bibfnamefont {H.}~\bibnamefont {Nakatsuji}},\
  }\href {\doibase 10.1063/1.2842068} {\bibfield  {journal} {\bibinfo
  {journal} {Journal of Chemical Physics}\ }\textbf {\bibinfo {volume} {128}},\
  \bibinfo {pages} {124103} (\bibinfo {year} {2008})}\BibitemShut {NoStop}%
\bibitem [{\citenamefont {Mironova}\ \emph {et~al.}(2015)\citenamefont
  {Mironova}, \citenamefont {Tupitsyn}, \citenamefont {Shabaev},\ and\
  \citenamefont {Plunien}}]{Mironova-2015}%
  \BibitemOpen
  \bibfield  {author} {\bibinfo {author} {\bibfnamefont {D.}~\bibnamefont
  {Mironova}}, \bibinfo {author} {\bibfnamefont {I.}~\bibnamefont {Tupitsyn}},
  \bibinfo {author} {\bibfnamefont {V.}~\bibnamefont {Shabaev}}, \ and\
  \bibinfo {author} {\bibfnamefont {G.}~\bibnamefont {Plunien}},\ }\href
  {\doibase https://doi.org/10.1016/j.chemphys.2015.01.003} {\bibfield
  {journal} {\bibinfo  {journal} {Chemical Physics}\ }\textbf {\bibinfo
  {volume} {449}},\ \bibinfo {pages} {10} (\bibinfo {year} {2015})}\BibitemShut
  {NoStop}%
\bibitem [{\citenamefont {Nogueira}\ and\ \citenamefont
  {Karr}(2023)}]{Nogueira-2023}%
  \BibitemOpen
  \bibfield  {author} {\bibinfo {author} {\bibfnamefont {H.~D.}\ \bibnamefont
  {Nogueira}}\ and\ \bibinfo {author} {\bibfnamefont {J.-P.}\ \bibnamefont
  {Karr}},\ }\href {\doibase 10.1103/PhysRevA.107.042817} {\bibfield  {journal}
  {\bibinfo  {journal} {Phys. Rev. A}\ }\textbf {\bibinfo {volume} {107}},\
  \bibinfo {pages} {042817} (\bibinfo {year} {2023})}\BibitemShut {NoStop}%
\bibitem [{\citenamefont {Alighanbari}\ \emph {et~al.}(2020)\citenamefont
  {Alighanbari}, \citenamefont {Giri}, \citenamefont {Constantin},
  \citenamefont {Korobov},\ and\ \citenamefont {Schiller}}]{Alighanbari2020}%
  \BibitemOpen
  \bibfield  {author} {\bibinfo {author} {\bibfnamefont {S.}~\bibnamefont
  {Alighanbari}}, \bibinfo {author} {\bibfnamefont {G.~S.}\ \bibnamefont
  {Giri}}, \bibinfo {author} {\bibfnamefont {F.~L.}\ \bibnamefont
  {Constantin}}, \bibinfo {author} {\bibfnamefont {V.~I.}\ \bibnamefont
  {Korobov}}, \ and\ \bibinfo {author} {\bibfnamefont {S.}~\bibnamefont
  {Schiller}},\ }\href {\doibase https://doi.org/10.1038/s41586-020-2261-5}
  {\bibfield  {journal} {\bibinfo  {journal} {Nature}\ }\textbf {\bibinfo
  {volume} {581}},\ \bibinfo {pages} {152} (\bibinfo {year}
  {2020})}\BibitemShut {NoStop}%
\bibitem [{\citenamefont {Patra}\ \emph {et~al.}(2020)\citenamefont {Patra},
  \citenamefont {Germann}, \citenamefont {Karr}, \citenamefont {Haidar},
  \citenamefont {Hilico}, \citenamefont {Korobov}, \citenamefont {Cozijn},
  \citenamefont {Eikema}, \citenamefont {Ubachs},\ and\ \citenamefont
  {Koelemeij}}]{Patra2020}%
  \BibitemOpen
  \bibfield  {author} {\bibinfo {author} {\bibfnamefont {S.}~\bibnamefont
  {Patra}}, \bibinfo {author} {\bibfnamefont {M.}~\bibnamefont {Germann}},
  \bibinfo {author} {\bibfnamefont {J.-P.}\ \bibnamefont {Karr}}, \bibinfo
  {author} {\bibfnamefont {M.}~\bibnamefont {Haidar}}, \bibinfo {author}
  {\bibfnamefont {L.}~\bibnamefont {Hilico}}, \bibinfo {author} {\bibfnamefont
  {V.~I.}\ \bibnamefont {Korobov}}, \bibinfo {author} {\bibfnamefont
  {F.~M.~J.}\ \bibnamefont {Cozijn}}, \bibinfo {author} {\bibfnamefont
  {K.~S.~E.}\ \bibnamefont {Eikema}}, \bibinfo {author} {\bibfnamefont
  {W.}~\bibnamefont {Ubachs}}, \ and\ \bibinfo {author} {\bibfnamefont
  {J.~C.~J.}\ \bibnamefont {Koelemeij}},\ }\href {\doibase
  10.1126/science.aba0453} {\bibfield  {journal} {\bibinfo  {journal}
  {Science}\ }\textbf {\bibinfo {volume} {369}},\ \bibinfo {pages} {1238}
  (\bibinfo {year} {2020})}\BibitemShut {NoStop}%
\bibitem [{\citenamefont {Kortunov}\ \emph {et~al.}(2021)\citenamefont
  {Kortunov}, \citenamefont {Alighanbari}, \citenamefont {Hansen},
  \citenamefont {Giri}, \citenamefont {Korobov},\ and\ \citenamefont
  {Schiller}}]{Kortunov2021}%
  \BibitemOpen
  \bibfield  {author} {\bibinfo {author} {\bibfnamefont {I.~V.}\ \bibnamefont
  {Kortunov}}, \bibinfo {author} {\bibfnamefont {S.}~\bibnamefont
  {Alighanbari}}, \bibinfo {author} {\bibfnamefont {M.~G.}\ \bibnamefont
  {Hansen}}, \bibinfo {author} {\bibfnamefont {G.~S.}\ \bibnamefont {Giri}},
  \bibinfo {author} {\bibfnamefont {V.~I.}\ \bibnamefont {Korobov}}, \ and\
  \bibinfo {author} {\bibfnamefont {S.}~\bibnamefont {Schiller}},\ }\href
  {\doibase https://doi.org/10.1038/s41567-020-01150-7} {\bibfield  {journal}
  {\bibinfo  {journal} {Nature Physics}\ }\textbf {\bibinfo {volume} {17}},\
  \bibinfo {pages} {569} (\bibinfo {year} {2021})}\BibitemShut {NoStop}%
\bibitem [{\citenamefont {Schiller}\ \emph
  {et~al.}(2014{\natexlab{a}})\citenamefont {Schiller}, \citenamefont
  {Bakalov},\ and\ \citenamefont {Korobov}}]{Schiller2014}%
  \BibitemOpen
  \bibfield  {author} {\bibinfo {author} {\bibfnamefont {S.}~\bibnamefont
  {Schiller}}, \bibinfo {author} {\bibfnamefont {D.}~\bibnamefont {Bakalov}}, \
  and\ \bibinfo {author} {\bibfnamefont {V.~I.}\ \bibnamefont {Korobov}},\
  }\href {\doibase 10.1103/PhysRevLett.113.023004} {\bibfield  {journal}
  {\bibinfo  {journal} {Phys. Rev. Lett.}\ }\textbf {\bibinfo {volume} {113}},\
  \bibinfo {pages} {023004} (\bibinfo {year} {2014}{\natexlab{a}})}\BibitemShut
  {NoStop}%
\bibitem [{\citenamefont {Schiller}\ \emph
  {et~al.}(2014{\natexlab{b}})\citenamefont {Schiller}, \citenamefont
  {Bakalov}, \citenamefont {Bekbaev},\ and\ \citenamefont
  {Korobov}}]{Schiller_P}%
  \BibitemOpen
  \bibfield  {author} {\bibinfo {author} {\bibfnamefont {S.}~\bibnamefont
  {Schiller}}, \bibinfo {author} {\bibfnamefont {D.}~\bibnamefont {Bakalov}},
  \bibinfo {author} {\bibfnamefont {A.~K.}\ \bibnamefont {Bekbaev}}, \ and\
  \bibinfo {author} {\bibfnamefont {V.~I.}\ \bibnamefont {Korobov}},\ }\href
  {\doibase 10.1103/PhysRevA.89.052521} {\bibfield  {journal} {\bibinfo
  {journal} {Phys. Rev. A}\ }\textbf {\bibinfo {volume} {89}},\ \bibinfo
  {pages} {052521} (\bibinfo {year} {2014}{\natexlab{b}})}\BibitemShut
  {NoStop}%
\bibitem [{\citenamefont {Johnson}\ \emph {et~al.}(1988)\citenamefont
  {Johnson}, \citenamefont {Blundell},\ and\ \citenamefont
  {Sapirstein}}]{Johnson_Bspline}%
  \BibitemOpen
  \bibfield  {author} {\bibinfo {author} {\bibfnamefont {W.~R.}\ \bibnamefont
  {Johnson}}, \bibinfo {author} {\bibfnamefont {S.~A.}\ \bibnamefont
  {Blundell}}, \ and\ \bibinfo {author} {\bibfnamefont {J.}~\bibnamefont
  {Sapirstein}},\ }\href {\doibase 10.1103/PhysRevA.37.307} {\bibfield
  {journal} {\bibinfo  {journal} {Phys. Rev. A}\ }\textbf {\bibinfo {volume}
  {37}},\ \bibinfo {pages} {307} (\bibinfo {year} {1988})}\BibitemShut
  {NoStop}%
\bibitem [{\citenamefont {Shabaev}\ \emph {et~al.}(2004)\citenamefont
  {Shabaev}, \citenamefont {Tupitsyn}, \citenamefont {Yerokhin}, \citenamefont
  {Plunien},\ and\ \citenamefont {Soff}}]{Shabaev_DKB}%
  \BibitemOpen
  \bibfield  {author} {\bibinfo {author} {\bibfnamefont {V.~M.}\ \bibnamefont
  {Shabaev}}, \bibinfo {author} {\bibfnamefont {I.~I.}\ \bibnamefont
  {Tupitsyn}}, \bibinfo {author} {\bibfnamefont {V.~A.}\ \bibnamefont
  {Yerokhin}}, \bibinfo {author} {\bibfnamefont {G.}~\bibnamefont {Plunien}}, \
  and\ \bibinfo {author} {\bibfnamefont {G.}~\bibnamefont {Soff}},\ }\href
  {\doibase 10.1103/PhysRevLett.93.130405} {\bibfield  {journal} {\bibinfo
  {journal} {Phys. Rev. Lett.}\ }\textbf {\bibinfo {volume} {93}},\ \bibinfo
  {pages} {130405} (\bibinfo {year} {2004})}\BibitemShut {NoStop}%
\bibitem [{\citenamefont {Rozenbaum}\ \emph {et~al.}(2014)\citenamefont
  {Rozenbaum}, \citenamefont {Glazov}, \citenamefont {Shabaev}, \citenamefont
  {Sosnova},\ and\ \citenamefont {Telnov}}]{Rozenbaum_ADKB}%
  \BibitemOpen
  \bibfield  {author} {\bibinfo {author} {\bibfnamefont {E.~B.}\ \bibnamefont
  {Rozenbaum}}, \bibinfo {author} {\bibfnamefont {D.~A.}\ \bibnamefont
  {Glazov}}, \bibinfo {author} {\bibfnamefont {V.~M.}\ \bibnamefont {Shabaev}},
  \bibinfo {author} {\bibfnamefont {K.~E.}\ \bibnamefont {Sosnova}}, \ and\
  \bibinfo {author} {\bibfnamefont {D.~A.}\ \bibnamefont {Telnov}},\ }\href
  {\doibase 10.1103/PhysRevA.89.012514} {\bibfield  {journal} {\bibinfo
  {journal} {Phys. Rev. A}\ }\textbf {\bibinfo {volume} {89}},\ \bibinfo
  {pages} {012514} (\bibinfo {year} {2014})}\BibitemShut {NoStop}%
\bibitem [{\citenamefont {Bakalov}\ \emph {et~al.}(2006)\citenamefont
  {Bakalov}, \citenamefont {Korobov},\ and\ \citenamefont
  {Schiller}}]{Bakalov2006}%
  \BibitemOpen
  \bibfield  {author} {\bibinfo {author} {\bibfnamefont {D.}~\bibnamefont
  {Bakalov}}, \bibinfo {author} {\bibfnamefont {V.~I.}\ \bibnamefont
  {Korobov}}, \ and\ \bibinfo {author} {\bibfnamefont {S.}~\bibnamefont
  {Schiller}},\ }\href {\doibase 10.1103/PhysRevLett.97.243001} {\bibfield
  {journal} {\bibinfo  {journal} {Phys. Rev. Lett.}\ }\textbf {\bibinfo
  {volume} {97}},\ \bibinfo {pages} {243001} (\bibinfo {year}
  {2006})}\BibitemShut {NoStop}%
\bibitem [{\citenamefont {Korobov}\ \emph {et~al.}(2014)\citenamefont
  {Korobov}, \citenamefont {Hilico},\ and\ \citenamefont {Karr}}]{Korobov2014}%
  \BibitemOpen
  \bibfield  {author} {\bibinfo {author} {\bibfnamefont {V.~I.}\ \bibnamefont
  {Korobov}}, \bibinfo {author} {\bibfnamefont {L.}~\bibnamefont {Hilico}}, \
  and\ \bibinfo {author} {\bibfnamefont {J.-P.}\ \bibnamefont {Karr}},\ }\href
  {\doibase 10.1103/PhysRevLett.112.103003} {\bibfield  {journal} {\bibinfo
  {journal} {Phys. Rev. Lett.}\ }\textbf {\bibinfo {volume} {112}},\ \bibinfo
  {pages} {103003} (\bibinfo {year} {2014})}\BibitemShut {NoStop}%
\bibitem [{\citenamefont {Korobov}\ \emph {et~al.}(2017)\citenamefont
  {Korobov}, \citenamefont {Hilico},\ and\ \citenamefont {Karr}}]{Korobov2017}%
  \BibitemOpen
  \bibfield  {author} {\bibinfo {author} {\bibfnamefont {V.~I.}\ \bibnamefont
  {Korobov}}, \bibinfo {author} {\bibfnamefont {L.}~\bibnamefont {Hilico}}, \
  and\ \bibinfo {author} {\bibfnamefont {J.-P.}\ \bibnamefont {Karr}},\ }\href
  {\doibase 10.1103/PhysRevLett.118.233001} {\bibfield  {journal} {\bibinfo
  {journal} {Phys. Rev. Lett.}\ }\textbf {\bibinfo {volume} {118}},\ \bibinfo
  {pages} {233001} (\bibinfo {year} {2017})}\BibitemShut {NoStop}%
\bibitem [{\citenamefont {Kotov}\ \emph {et~al.}(2020)\citenamefont {Kotov},
  \citenamefont {Glazov}, \citenamefont {Malyshev}, \citenamefont
  {Vladimirova}, \citenamefont {Shabaev},\ and\ \citenamefont
  {Plunien}}]{Kotov2020}%
  \BibitemOpen
  \bibfield  {author} {\bibinfo {author} {\bibfnamefont {A.~A.}\ \bibnamefont
  {Kotov}}, \bibinfo {author} {\bibfnamefont {D.~A.}\ \bibnamefont {Glazov}},
  \bibinfo {author} {\bibfnamefont {A.~V.}\ \bibnamefont {Malyshev}}, \bibinfo
  {author} {\bibfnamefont {A.~V.}\ \bibnamefont {Vladimirova}}, \bibinfo
  {author} {\bibfnamefont {V.~M.}\ \bibnamefont {Shabaev}}, \ and\ \bibinfo
  {author} {\bibfnamefont {G.}~\bibnamefont {Plunien}},\ }\href {\doibase
  https://doi.org/10.1002/xrs.3064} {\bibfield  {journal} {\bibinfo  {journal}
  {X-Ray Spectrometry}\ }\textbf {\bibinfo {volume} {49}},\ \bibinfo {pages}
  {110} (\bibinfo {year} {2020})}\BibitemShut {NoStop}%
\bibitem [{\citenamefont {Kotov}\ \emph {et~al.}(2021)\citenamefont {Kotov},
  \citenamefont {Glazov}, \citenamefont {Shabaev},\ and\ \citenamefont
  {Plunien}}]{Kotov2021}%
  \BibitemOpen
  \bibfield  {author} {\bibinfo {author} {\bibfnamefont {A.~A.}\ \bibnamefont
  {Kotov}}, \bibinfo {author} {\bibfnamefont {D.~A.}\ \bibnamefont {Glazov}},
  \bibinfo {author} {\bibfnamefont {V.~M.}\ \bibnamefont {Shabaev}}, \ and\
  \bibinfo {author} {\bibfnamefont {G.}~\bibnamefont {Plunien}},\ }\href
  {\doibase 10.3390/atoms9030044} {\bibfield  {journal} {\bibinfo  {journal}
  {Atoms}\ }\textbf {\bibinfo {volume} {9}} (\bibinfo {year} {2021}),\
  10.3390/atoms9030044}\BibitemShut {NoStop}%
\bibitem [{\citenamefont {Kotov}\ \emph {et~al.}(2022)\citenamefont {Kotov},
  \citenamefont {Glazov}, \citenamefont {Malyshev}, \citenamefont {Shabaev},\
  and\ \citenamefont {Plunien}}]{Kotov2022}%
  \BibitemOpen
  \bibfield  {author} {\bibinfo {author} {\bibfnamefont {A.~A.}\ \bibnamefont
  {Kotov}}, \bibinfo {author} {\bibfnamefont {D.~A.}\ \bibnamefont {Glazov}},
  \bibinfo {author} {\bibfnamefont {A.~V.}\ \bibnamefont {Malyshev}}, \bibinfo
  {author} {\bibfnamefont {V.~M.}\ \bibnamefont {Shabaev}}, \ and\ \bibinfo
  {author} {\bibfnamefont {G.}~\bibnamefont {Plunien}},\ }\href {\doibase
  10.3390/atoms10040145} {\bibfield  {journal} {\bibinfo  {journal} {Atoms}\
  }\textbf {\bibinfo {volume} {10}} (\bibinfo {year} {2022}),\
  10.3390/atoms10040145}\BibitemShut {NoStop}%
\bibitem [{\citenamefont {Sapirstein}\ and\ \citenamefont
  {Johnson}(1996)}]{Sapirstein_1996}%
  \BibitemOpen
  \bibfield  {author} {\bibinfo {author} {\bibfnamefont {J.}~\bibnamefont
  {Sapirstein}}\ and\ \bibinfo {author} {\bibfnamefont {W.~R.}\ \bibnamefont
  {Johnson}},\ }\href {\doibase 10.1088/0953-4075/29/22/005} {\bibfield
  {journal} {\bibinfo  {journal} {Journal of Physics B: Atomic, Molecular and
  Optical Physics}\ }\textbf {\bibinfo {volume} {29}},\ \bibinfo {pages} {5213}
  (\bibinfo {year} {1996})}\BibitemShut {NoStop}%
\bibitem [{\citenamefont {Carrington}\ \emph {et~al.}(1989)\citenamefont
  {Carrington}, \citenamefont {McNab},\ and\ \citenamefont
  {Montgomerie}}]{Carrington_1989}%
  \BibitemOpen
  \bibfield  {author} {\bibinfo {author} {\bibfnamefont {A.}~\bibnamefont
  {Carrington}}, \bibinfo {author} {\bibfnamefont {I.~R.}\ \bibnamefont
  {McNab}}, \ and\ \bibinfo {author} {\bibfnamefont {C.~A.}\ \bibnamefont
  {Montgomerie}},\ }\href {\doibase 10.1088/0953-4075/22/22/006} {\bibfield
  {journal} {\bibinfo  {journal} {Journal of Physics B: Atomic, Molecular and
  Optical Physics}\ }\textbf {\bibinfo {volume} {22}},\ \bibinfo {pages} {3551}
  (\bibinfo {year} {1989})}\BibitemShut {NoStop}%
\bibitem [{\citenamefont {Korobov}\ and\ \citenamefont
  {Tsogbayar}(2007)}]{Korobov_2007}%
  \BibitemOpen
  \bibfield  {author} {\bibinfo {author} {\bibfnamefont {V.~I.}\ \bibnamefont
  {Korobov}}\ and\ \bibinfo {author} {\bibfnamefont {T.}~\bibnamefont
  {Tsogbayar}},\ }\href {\doibase 10.1088/0953-4075/40/13/011} {\bibfield
  {journal} {\bibinfo  {journal} {Journal of Physics B: Atomic, Molecular and
  Optical Physics}\ }\textbf {\bibinfo {volume} {40}},\ \bibinfo {pages} {2661}
  (\bibinfo {year} {2007})}\BibitemShut {NoStop}%
\bibitem [{\citenamefont {Tsogbayar}\ and\ \citenamefont
  {Korobov}(2006)}]{Korobov_2006}%
  \BibitemOpen
  \bibfield  {author} {\bibinfo {author} {\bibfnamefont {T.}~\bibnamefont
  {Tsogbayar}}\ and\ \bibinfo {author} {\bibfnamefont {V.~I.}\ \bibnamefont
  {Korobov}},\ }\href {\doibase 10.1063/1.2209694} {\bibfield  {journal}
  {\bibinfo  {journal} {The Journal of Chemical Physics}\ }\textbf {\bibinfo
  {volume} {125}} (\bibinfo {year} {2006}),\ 10.1063/1.2209694}\BibitemShut
  {NoStop}%
\bibitem [{\citenamefont {Eyring}(1964)}]{QC-Eyring}%
  \BibitemOpen
  \bibfield  {author} {\bibinfo {author} {\bibfnamefont {H.}~\bibnamefont
  {Eyring}},\ }\href@noop {} {\emph {\bibinfo {title} {Quantum chemistry}}},\
  \bibinfo {edition} {12th}\ ed.\ (\bibinfo  {publisher} {John Wiley $\&$
  Sons},\ \bibinfo {address} {New York},\ \bibinfo {year} {1964})\BibitemShut
  {NoStop}%
\bibitem [{\citenamefont {Pauling}(1933)}]{Pauling1933}%
  \BibitemOpen
  \bibfield  {author} {\bibinfo {author} {\bibfnamefont {L.}~\bibnamefont
  {Pauling}},\ }\href {\doibase 10.1063/1.1749219} {\bibfield  {journal}
  {\bibinfo  {journal} {The Journal of Chemical Physics}\ }\textbf {\bibinfo
  {volume} {1}},\ \bibinfo {pages} {56} (\bibinfo {year} {1933})}\BibitemShut
  {NoStop}%
\bibitem [{\citenamefont {Janev}\ \emph {et~al.}(1997)\citenamefont {Janev},
  \citenamefont {Pop-Jordanov},\ and\ \citenamefont {Solov'ev}}]{Janev_1997}%
  \BibitemOpen
  \bibfield  {author} {\bibinfo {author} {\bibfnamefont {R.~K.}\ \bibnamefont
  {Janev}}, \bibinfo {author} {\bibfnamefont {J.}~\bibnamefont {Pop-Jordanov}},
  \ and\ \bibinfo {author} {\bibfnamefont {E.~A.}\ \bibnamefont {Solov'ev}},\
  }\href {\doibase 10.1088/0953-4075/30/10/005} {\bibfield  {journal} {\bibinfo
   {journal} {Journal of Physics B: Atomic, Molecular and Optical Physics}\
  }\textbf {\bibinfo {volume} {30}},\ \bibinfo {pages} {L353} (\bibinfo {year}
  {1997})}\BibitemShut {NoStop}%
\bibitem [{\citenamefont {Baxter}\ and\ \citenamefont
  {Kirchner}(2016)}]{Baxter2016}%
  \BibitemOpen
  \bibfield  {author} {\bibinfo {author} {\bibfnamefont {M.}~\bibnamefont
  {Baxter}}\ and\ \bibinfo {author} {\bibfnamefont {T.}~\bibnamefont
  {Kirchner}},\ }\href {\doibase 10.1103/PhysRevA.93.012502} {\bibfield
  {journal} {\bibinfo  {journal} {Phys. Rev. A}\ }\textbf {\bibinfo {volume}
  {93}},\ \bibinfo {pages} {012502} (\bibinfo {year} {2016})}\BibitemShut
  {NoStop}%
\bibitem [{\citenamefont {Minami}\ \emph {et~al.}(2007)\citenamefont {Minami},
  \citenamefont {Pindzola}, \citenamefont {Lee},\ and\ \citenamefont
  {Schultz}}]{Minami_2007}%
  \BibitemOpen
  \bibfield  {author} {\bibinfo {author} {\bibfnamefont {T.}~\bibnamefont
  {Minami}}, \bibinfo {author} {\bibfnamefont {M.~S.}\ \bibnamefont
  {Pindzola}}, \bibinfo {author} {\bibfnamefont {T.-G.}\ \bibnamefont {Lee}}, \
  and\ \bibinfo {author} {\bibfnamefont {D.~R.}\ \bibnamefont {Schultz}},\
  }\href {\doibase 10.1088/0953-4075/40/18/005} {\bibfield  {journal} {\bibinfo
   {journal} {Journal of Physics B: Atomic, Molecular and Optical Physics}\
  }\textbf {\bibinfo {volume} {40}},\ \bibinfo {pages} {3629} (\bibinfo {year}
  {2007})}\BibitemShut {NoStop}%
\bibitem [{\citenamefont {Faulkner}\ \emph {et~al.}(2019)\citenamefont
  {Faulkner}, \citenamefont {Abdurakhmanov}, \citenamefont {Alladustov},
  \citenamefont {Kadyrov},\ and\ \citenamefont {Bray}}]{Faulkner_2019}%
  \BibitemOpen
  \bibfield  {author} {\bibinfo {author} {\bibfnamefont {J.}~\bibnamefont
  {Faulkner}}, \bibinfo {author} {\bibfnamefont {I.~B.}\ \bibnamefont
  {Abdurakhmanov}}, \bibinfo {author} {\bibfnamefont {S.~U.}\ \bibnamefont
  {Alladustov}}, \bibinfo {author} {\bibfnamefont {A.~S.}\ \bibnamefont
  {Kadyrov}}, \ and\ \bibinfo {author} {\bibfnamefont {I.}~\bibnamefont
  {Bray}},\ }\href {\doibase 10.1088/1361-6587/ab2e7a} {\bibfield  {journal}
  {\bibinfo  {journal} {Plasma Physics and Controlled Fusion}\ }\textbf
  {\bibinfo {volume} {61}},\ \bibinfo {pages} {095005} (\bibinfo {year}
  {2019})}\BibitemShut {NoStop}%
\bibitem [{\citenamefont {Angeli}\ and\ \citenamefont
  {Marinova}(2013)}]{angeli2013table}%
  \BibitemOpen
  \bibfield  {author} {\bibinfo {author} {\bibfnamefont {I.}~\bibnamefont
  {Angeli}}\ and\ \bibinfo {author} {\bibfnamefont {K.~P.}\ \bibnamefont
  {Marinova}},\ }\href@noop {} {\bibfield  {journal} {\bibinfo  {journal}
  {Atomic Data and Nuclear Data Tables}\ }\textbf {\bibinfo {volume} {99}},\
  \bibinfo {pages} {69} (\bibinfo {year} {2013})}\BibitemShut {NoStop}%
\bibitem [{\citenamefont {Furlanetto}\ and\ \citenamefont
  {Furlanetto}(2007)}]{Furlanetto2007}%
  \BibitemOpen
  \bibfield  {author} {\bibinfo {author} {\bibfnamefont {S.~R.}\ \bibnamefont
  {Furlanetto}}\ and\ \bibinfo {author} {\bibfnamefont {M.~R.}\ \bibnamefont
  {Furlanetto}},\ }\href {\doibase 10.1111/j.1365-2966.2007.11921.x} {\bibfield
   {journal} {\bibinfo  {journal} {Monthly Notices of the Royal Astronomical
  Society}\ }\textbf {\bibinfo {volume} {379}},\ \bibinfo {pages} {130}
  (\bibinfo {year} {2007})}\BibitemShut {NoStop}%
\bibitem [{\citenamefont {Ravi}\ \emph {et~al.}(2021)\citenamefont {Ravi},
  \citenamefont {Mukherjee}, \citenamefont {Mukherjee}, \citenamefont
  {Adhikari}, \citenamefont {Sathyamurthy},\ and\ \citenamefont
  {Baer}}]{Ravi2021}%
  \BibitemOpen
  \bibfield  {author} {\bibinfo {author} {\bibfnamefont {S.}~\bibnamefont
  {Ravi}}, \bibinfo {author} {\bibfnamefont {S.}~\bibnamefont {Mukherjee}},
  \bibinfo {author} {\bibfnamefont {B.}~\bibnamefont {Mukherjee}}, \bibinfo
  {author} {\bibfnamefont {S.}~\bibnamefont {Adhikari}}, \bibinfo {author}
  {\bibfnamefont {N.}~\bibnamefont {Sathyamurthy}}, \ and\ \bibinfo {author}
  {\bibfnamefont {M.}~\bibnamefont {Baer}},\ }\href {\doibase
  10.1080/00268976.2020.1811907} {\bibfield  {journal} {\bibinfo  {journal}
  {Molecular Physics}\ }\textbf {\bibinfo {volume} {119}},\ \bibinfo {pages}
  {e1811907} (\bibinfo {year} {2021})}\BibitemShut {NoStop}%
\end{thebibliography}%

\end{document}